# Active Learning in Physics: From 101, to Progress, and Perspective

Yongcheng Ding[1,2], José D. Martín-Guerrero[3,4], Yolanda Vives-Gilabert[3], Xi Chen[1,5]

[1] *Department of Physical Chemistry, University of the Basque Country UPV/EHU, Apartado 644, 48080 Bilbao, Spain*

[2] *International Center of Quantum Artificial Intelligence for Science and Technology (QuArtist) and Department of Physics, Shanghai University, 200444 Shanghai, People's Republic of China*

[3] *Intelligent Data Analysis Laboratory (IDAL), Department of Electronic Engineering, ETSE-UV, Universitat de València. Avgda. Universitat, s/n. 46100 Burjassot (Valencia), Spain*

[4] *Valencian Graduate School and Research Network of Artificial Intelligence (ValgrAI), Valencia, Spain*

[5] *EHU Quantum Center, University of the Basque Country UPV/EHU, Barrio Sarriena, s/n, 48940 Leioa, Spain*

**Abstract**

Active Learning (AL) is a family of machine learning (ML) algorithms that predates the current era of artificial intelligence. Unlike traditional approaches that require labeled samples for training, AL iteratively selects unlabeled samples to be annotated by an expert. This protocol aims to prioritize the most informative samples, leading to improved model performance compared to training with all labeled samples. In recent years, AL has gained increasing attention, particularly in the field of physics. This paper presents a comprehensive and accessible introduction to the theory of AL reviewing the latest advancements across various domains. Additionally, we explore the potential integration of AL with quantum ML, envisioning a synergistic fusion of these two fields rather than viewing AL as a mere extension of classical ML into the quantum realm.

## Background

With the remarkable progress of machine learning (ML), the notion of "Artificial Intelligence for science" has become deeply ingrained in the minds of computer scientists. We are constantly enticed by the idea of tackling challenging physics problems with an omniscient model that seems to exist in our imagination. However, a significant obstacle stands in our path, and it primarily consists in the expensive labeling process in supervised learning. Labels are indeed expensive by any degree. For instance, in fields like computer vision or language processing, training sets often require a multitude of figures to be painstakingly labeled, typically done manually. This has led to a demand for hiring individuals solely dedicated to this task or even the infamous reCAPTCHA tests we have endured for years. As physicists, our situation is even more daunting. The labels for our samples are usually derived from experiments and data analysis. This means we rely on highly skilled human annotators, research funding, and considerable time investments for both numerical simulations and laboratory experiments. To overcome this challenge, active learning (AL) emerges as a family of ML algorithms designed to alleviate the labeling cost, delivering astonishing results in various industries. The underlying concept is based on the hypothesis that not every sample in the dataset needs to be labeled to train a supervised ML model effectively. Instead, only a small subset of samples, specifically the most informative ones, can be labeled, resulting in a highly efficient training process with a significantly reduced training set. This naturally raises the question: how do we determine which samples are the most informative ones? Although we are initially uncertain about the labels of these samples, we can rely on the ML model itself to assess their informativeness. The model can evaluate each sample's potential for providing valuable information, and the specific criteria for making such judgments will be discussed later. With this workflow in mind, we can embark on the AL journey to train our models. It begins with initializing the model using a small set of labeled samples. Through an iterative process, we select the most informative unlabeled sample from the pool, present it to an oracle (e.g., a human annotator) for labeling, expand the training set for supervised learning, and update the model's parameters [Settles2009, Sverchkov2017, Tuia2011]. This process continues until convergence is achieved or until the cost of acquiring another label becomes prohibitive. By embracing AL, we can surmount the challenges posed by expensive labeling and unleash the full potential of our ML models in the realm of physics and beyond.

Delving into the details of the strategies used to select unlabeled samples for querying. It is important to note that there are numerous methods available for this task. However, in this discussion, we will focus on beginner-level strategies and reserve the exploration of advanced approaches for specific scenarios. By understanding these fundamental strategies, we can gain valuable insights into the initial steps of AL and lay a solid foundation for further exploration. The starting hypothesis is that we have a ML model characterized by its parameters θ, which is either initialized by randomization, or trained by a set of labeled samples $X = \{x_i, y_i\}_{i=1}^{l}$ with $x_i$ be vectors as inputs and $y_i$ be labels. We also have a pool of unlabeled samples $U = \{x_i\}_{i=l+1}^{l+u}$, which is much larger than the size of the training set.

**Uncertainty sampling**

The most basic strategy in AL revolves around the belief that the sample with the greatest uncertainty will provide the most valuable information once labeled [Zhu2010]. To implement this strategy, we calculate the uncertainty of all unlabeled samples in the pool based on the model's estimation and select the candidate with the highest uncertainty for labeling by a human annotator. The key question here is: how do we quantify uncertainty using a ML model? One intuitive approach is known as the "least confidence" method, where the model estimates the probabilities associated with each potential label for a given sample. We query the sample according to the criteria

$$x_{LC} = \text{argmax}_x[1 - P_\theta(\hat{y}|x)],$$

$$\hat{y} = \text{argmax}_y[P_\theta(y|x)],$$

where $\hat{y}$ is the class with the most confidence estimated by the model θ. Once the sample $x_{LC}$ is selected and labeled by the human annotator, we transfer it from the pool $U$ to the training set $X$, and update the parameters θ of the model.

**Margin sampling**
One immediately notices that the least confidence only considers the information of the most probable class $\widehat{y_1}$, i.e., information of other labels is lost. By taking the second most probable class $\widehat{y_2}$ into account, one has the margin sampling as a more informative strategy

$$x_M = \text{argmin}_x[P_\theta(\widehat{y_1}|x) - P_\theta(\widehat{y_2}|x)],$$

which stands for the experience that it is harder for one to classify ambiguous samples separated by smaller margin. By querying the sample with the minimal margin, one introduces more information to the training set to discriminate among all classes [Zhu2010].

**Entropy sampling**
According to Shannon's theory, one can characterize the amount of information by entropy after excluding miscellaneous information, and averaging all uncertainties [Zhu2010]. Thus, one can employ the information entropy that takes information from all classes as

$$x_E = \mathrm{argmax}_x \left[ -\sum_i P_\theta(\hat{y}_i|x) \log P_\theta(\hat{y}_i|x) \right].$$

For uncertainty sampling, these three strategies exhibit different behaviors across multiple class problems and various pools and training sets, particularly when the samples are imbalanced. However, it can be proven that they are equivalent when dealing with binary classification problems. To illustrate this point, let's consider a default support-vector-machine that seeks to separate the parameter space for binary classification. All these strategies would converge to selecting the sample that is closest to the decision boundary. This demonstrates the model-free nature of AL.

**Query by committee**
Instead of relying on uncertainty sampling with a single model, the query by committee approach involves using a committee of models, assuming they are available for the problem. In this strategy, the candidate is selected based on consensus rather than uncertainty. Similar to entropy sampling, the committee utilizes voting entropy to determine the minimum consensus as

$$x_{VE} = \mathrm{argmax}_x \left[ -\sum_i \frac{V(y_i)}{C} \log \frac{V(y_i)}{C} \right],$$

where $V(y_i)$ is the vote results for label $y\_i$ from the committee of $C$ models. One can also use the Kullback-Leilber divergence as the measure of minimal consensus as

$$x_{KL} = \mathrm{argmax}_x \left[ \frac{1}{C} \sum_{i=1}^{C} D(P|P_C) \right],$$

$$D(P_{\theta^{(i)}}|P_C) = \sum_i P_{\theta^{(i)}}(y_i|x) \log \frac{P_{\theta^{(i)}}(y_i|x)}{P_C(y_i|x)},$$

where $\theta^{(i)}$ is the $i$-th model in the committee and $P_C(y_i|x) = \frac{1}{C}\sum_{c=1}^{C} P_{\theta^{(i)}}(y_i|x)$ characterizes the probability of making an agreement on that $y_i$ is the correct label. The Kullback-Leibler divergence serves as an information-theoretic metric to quantify the disparity between two probability distributions. Hence, this measure of discrepancy prioritizes the most informative query by assessing the highest average

distinction between the label distributions of any individual committee member and the consensus [Seung1992].

Based on the experience, query by committee usually outperforms uncertainty sampling by accuracy if the maximum size of the training set is bounded at the cost of extra computational resources required by the committee. Meanwhile, we have to emphasize that the models in the committee should be suitable for the task. Otherwise, near-to-random predictions by models degenerate the strategy to random sampling, as the baseline for benchmarking the validity of AL, or even worse, a committee of wrong models might result in prediction that is overtaken by blind-guessing.

## Variants of Active Learning

According to the definition, an AL algorithm consists of a sampling strategy, iteratively updated training set and model, as well as an oracle. The term "active" gives rise to different variants of ML algorithms and has sparked impressive research in the field. While this review does not aim to engage in a debate about terminology, it is important to include and differentiate these variants to enhance our understanding of the various protocols.

**Reinforcement Learning**
An agent in reinforcement learning (RL) can take actions once it observes states based on the policy. It is quite natural that this algorithm is employed for solving quantum control problems such as state preparation or quantum gate design. Meanwhile, RL also allows the agent to actively propose quantum experiments to explore intricate quantum phenomena. To be more specific, researchers introduce an autonomous learning model that learns to design these experiments without relying on prior knowledge or human intuition [Melnikov2018]. The model not only outperformed previous approaches in terms of efficiency but also discovered new experimental techniques. By using the projective simulation model, they successfully designed photonic quantum experiments generating high-dimensional entangled multiphoton states, a topic of significant interest in modern quantum experiments. The AI system exhibited creativity by autonomously rediscovering experimental techniques that are now considered standard in contemporary quantum optical experiments. By showcasing the capability of ML, this study revealed how it has the potential to revolutionize the generation of experiments. Additionally, it emphasized the significant role played by intelligent machines in advancing scientific research through their assistance in experimental design, akin to other ML algorithms [Krenn2016].

**Semisupervised learning**
Semisupervised learning follows a similar protocol to AL, starting with a small training set that is iteratively updated as samples are selected from the pool and labeled. It shares the goal of reducing labeling costs, but there is a crucial difference: instead of querying the most uncertain samples, it selects samples with the highest confidence in their labels based on the model's estimation or cluster algorithms. These labeled samples are then incorporated into the training set. This approach allows semisupervised learning to train a model at a lower cost compared to AL, as it eliminates the need for external oracles, typically, human annotators. For example, regarding evaporative cooling experiments, researchers developed a ML-based approach to optimize experimental control [Wu2020]. The method utilizes neural networks to learn the relationship between control parameters and the desired control goal, enabling the determination of optimal control parameters. A key challenge in this approach is the limited availability of labeled data from experiments. To address this challenge, the authors employed semisupervised learning (according to our definition), which allows for overcoming data scarcity. The effectiveness of their scheme was demonstrated through the control of cooling experiments in cold atoms. Initially, the method was tested using simulated data, followed by its application to real experiments. The results showcased that the proposed method achieved superior performance within a few hundred experimental runs. Notably, their approach does not require prior knowledge of the specific experimental system and holds applicability across different systems for experimental control.

## Progress

In recent years, AL has found applications in the field of physical science, particularly in material science and chemistry. It is worth noting that the connection between AL and quantum physics was initially proposed for accelerating AL through the principles of quantum mechanics [Paparo2014]. However, it is important to clarify that the definition of AL is rooted in RL, as we discussed in the previous section. Another interesting development is the application of quantum support vector machines for sampling against adversarial attacks, demonstrating the potential of quantum AL algorithms with polylogarithmic complexity [Casares2020]. While the utilization of quantum algorithms to enhance AL is a compelling topic, our focus in this paper is on the applications of AL in the realm of physical science. Physicists often face challenges in optimizing classical algorithms with their physical insights, making it more feasible for them to leverage well-established theories from computer science for their research endeavors.

**Quantum Information**
Before delving into the practical applications of AL in the field of physical science, it is important to first introduce quantum information retrieval with AL. This particular area of research is well-suited for AL with the precise definition as presented in this paper. Moreover, it serves as a natural playground for AL due to the inherent cost of labeling, which can be characterized as fidelity loss resulting from quantum measurements. By exploring quantum information retrieval with AL, we can gain valuable insights and establish a solid foundation for further applications in other domains within the realm of physical science. Problems are proposed as binary and multinomial classification [Ding2020, Ding2021], which are easy to be solved by support vector machines as toy models. The only difference between the quantum version and the classical version is that the label of each sample is determined by the amplitudes on eigenbases, being encoded in two-level or multi-level quantum systems as quantum information for extraction via measurements with apparatuses. Although the problems are very simple, these allows one to benchmark AL of common strategies with random sampling as the baseline, demonstrating the feasibility by the rate estimation over 90% with only 5% samples labeled. Moreover, it offers a tool to characterize and compare the information extraction by von Neumann measurement and weak measurement. The von Neumann measurement extracts the information with higher signal-to-noise-ratio, collapses the wave function, and induce significant fidelity loss. In the contrary, weak measurement extracts less effective information with less perturbation on the wave function. By bounding the fidelity loss, AL trains the model with different information extraction protocol, and compare them by accuracy rate as figure-of-merit, implying the trade-off between copies, threshold of fidelity loss, and measurement strength. In the realm of more practical applications, an active-learning-inspired algorithm has been proposed to estimate the state fidelity or gate fidelity by optimizing the settings of quantum measurements [Zhu2022]. This approach can be integrated into the training of variational quantum algorithms, enhancing their performance. Additionally, standard AL has made significant progress in the field of quantum state tomography [Lange2023], which is considered a milestone in quantum information retrieval. In this context, the sampling strategy involves the use of query by committee, while the ML model is based on a restricted Boltzmann machine. The success of this approach is evident in its adaptive reconstruction of the ground state of the XXZ model and kinetically constrained spin chain model, highlighting the power of AL in addressing complex problems in many-body physics.

**High energy physics**
The introduction of AL techniques holds significant potential in the field of high energy physics, encompassing disciplines such as particle physics, nuclear physics, and plasma physics (high-energy-density physics for the latter two). The experimental cost for studies in this field, either laboratory or numerical, is usually enormous, e.g., the accelerators, colliders, as well as the supercomputers. Meanwhile, one can train a ML model with AL to reduce the cost. For example, query by committee and its variant with dropout has been proposed to constrain the parameters of supersymmetry model in 19 dimensions [Caron2019], identify uncertain regions, and steering new searches. For going beyond the Standard Model and Higgs bosons, an AL approach is presented for predicting the compatibility between new physics theories and existing experimental data obtained from particle colliders [Rocamonde2022]. This approach has achieved over 90% rate estimation using significantly fewer computing resources, accounting for less than 10% of the computational requirements of earlier methods, enabling the examination of previously untestable models and facilitates large-scale evaluations of new physics theories. In nuclear physics, margin sampling is applied to map out the thermodynamics stability of the quantum chromodynamics equation [Mroczek2023], which is developed by the Beam Energy Scan Theory collaboration. It utilizes a nonuniversal linear mapping of 3D Ising model variables onto the phase diagram, which involves four free parameters. Within the resulting four-dimensional parameter space, certain combinations lead to unstable or acausal realizations of the equation. The AL framework focuses on the most crucial regions in the input parameter space. This approach efficiently identifies and eliminates unphysical equation of state instances with high accuracy. For soft matter physics, query by committee strategy with an ensemble of neural networks efficiently simulates the near-equilibrium plasma flows at multiscale [Diaw2020]. To overcome the computational cost of gathering information from molecular dynamics with kinetic theory, AL is employed to train neural networks using a small randomly sampled subset of the parameter space. The method is applied to investigate a plasma interfacial mixing problem relevant to warm dense matter, demonstrating significant computational efficiency compared to the full kinetic-molecular-dynamics approach. The results indicate that this approach enables the exploration of Coulomb coupling physics across a wide range of temperatures, and densities that are currently beyond the reach of existing theoretical models.

**Condensed Matter Physics**

AL has gained significant traction in this field, particularly due to its relevance in computational physics and material physics experiments. As the dimension of the Hilbert space expands exponentially with the system's size, computational resources become increasingly vital for Monte Carlo algorithms, density functional theory, and molecular dynamics. Moreover, sample preparation in material science is both costly and time-consuming. A subfield that benefits from AL is the computational approach to many-body physics. In addition to employing state tomography, the technique of query by committee is proposed to economically label data by utilizing an ensemble of neural networks to effectively fit the multidimensional function [Yao2020]. Researchers outline the overall protocol of their fitting scheme and provide a detailed procedure for computing physical observables using the fitted functions. To demonstrate the effectiveness of the method, they present two examples: the quantum three-body problem in atomic physics and the calculation of anomalous Hall conductivity in condensed matter physics, yielding satisfactory outcomes in both cases. AL has shown promise in estimating phase boundaries and predicting exotic phenomena such as quantum phase transitions [Ding2022]. In the context of the antiferromagnetic Ising model on a triangular lattice under a transverse field, support vector machines, employing Gaussian kernel and uncertainty sampling, successfully predict the boundaries between the ordered phase, Kosterlitz-Thouless phase, and paramagnetic phases. By leveraging only a small number of data points obtained from the quantum Monte Carlo algorithm, the predicted phase boundaries closely resemble the analytical solutions, demonstrating a remarkable level of similarity. In material science, the well-celebrated study has achieved the uniformly accurate interatomic potentials for materials simulation with deep potential generator as the variant of AL [Zhang2019], which comprises three key components: exploration, generation of precise reference data, and training. The application of this procedure to sample systems such as Al, Mg, and Al-Mg alloys, showcases the capability of the model to create uniformly accurate potential energy surface models using a minimal amount of reference data. Furthermore, by combining AL and element embedding approach, the formation energies of oxygen vacancy layers, lattice parameters, and their statistical correlations in infinite-layer and perovskite oxides across the periodic table have been explored [Sahinovic2021]. By considering the Kullback-Leibler divergence, they show that the neural networks can predict these observables with 30% of the data in the pool, achieving high precision, while the AL algorithm compose the training set without human knowledge of chemistry. AL has estimated the material property curves and surfaces [Tian2021], focusing on the challenge of determining material properties in relation to independent variables, which typically involves time-consuming experiments or calculations. They have compared various sampling strategies based on directed exploration using a Kriging-based model across different materials problems of varying complexity. The atomic cluster models [Lysogorskiy2023] has also been studied: one based on the D-optimality criterion and the other utilizing ensemble learning. The extrapolation grade indicator facilitates active exploration of new structures, potentially leading to the automated discovery of rare-event configurations. The study also demonstrates the applicability of AL in exploring local atomic environments using large-scale molecular dynamics simulations.

## Perspective: Quantum Machine Learning

Here, we present our perspective on the extension of AL to the emerging field of quantum ML. The term "quantum machine learning" encompasses ML using quantum devices or quantum computers, including the utilization of artificial neural networks to accelerate ML models [Biamonte2017]. Quantum ML holds great promise for enhancing the efficiency of ML processes. For example, it can be employed to analyze outputs from quantum physics experiments as a post-processing step, capitalizing on the availability of quantum states as inputs. Additionally, leveraging this paradigm to address classical problems and explore potential quantum advantages becomes intriguing when classical data can be encoded into quantum systems as quantum information. In both scenarios, AL plays a crucial role, as quantum states are fragile, being inevitable to collapse or perturbate when extracting information for labeling; besides, they are non-replicable due to the fundamental principles of quantum mechanics. Generating sufficient copies of quantum states for labeling or training necessitates repetitive data encoding or experimentation and storage in quantum memories, or on-demand creation upon query. Therefore, any technique that reduces the number of samples and labels required for training a quantum device or quantum neural network represents the silver bullet to practical quantum ML. To our best knowledge, the pioneering quantum AL is performed in a programmable photonic quantum processor [Ding2023]. In this study, two AL-enabled variational quantum classifiers were designed and implemented. A programmable free-space photonic quantum processor was employed to execute these classifiers, comparing their performance with and without the query-by-committee strategy. The results demonstrated the significant advantages of AL in quantum ML, with labeling efforts reduced by up to 85% and computational efforts reduced by more than 90% in a data classification task. These findings highlight its potential and effectiveness in quantum ML, inspiring further applications in large-scale settings to enhance training efficiency and explore practical quantum advantages in quantum physics and real-world applications.

As the field of quantum ML continues to evolve, the development of specialized strategies becomes crucial for harnessing the full potential of AL in quantum systems.

**Expected model change**

We propose introducing an additional strategy in AL called "expected model change". This strategy focuses on selecting samples that are expected to induce the greatest change in the model, provided we have knowledge of their labels. A specific example of this strategy is the "expected gradient length," which can be applied to any learning problem that utilizes gradient-based training. Consequently, it can be effectively combined with a variational quantum circuit as the learning model, which is typically trained by minimizing a loss function based on the expectation values of observables using gradient methods. In other words, the AL process aims to query a sample, denoted as $x$, which, when labeled and added to the training set $X$, will lead to the largest magnitude change in the training gradient. Let $\nabla l_\theta(X)$ be the gradient of the loss function with respect to the gate parameter, and $\nabla l_\theta(X \cup \{x, y\})$ be the updated gradient once the labeled sample $\{x, y\}$ is transferred to the training set. We select the sample the with maximal gradient length as

$$x_{EGL} = \operatorname{argmax}_x \sum_i P_\theta(y_i|x) \operatorname{norm}[\nabla l_\theta(X \cup \{x, y_i\})],$$

where $\operatorname{norm}$ calculates the Euclidean norm of each gradient vector. The criteria can be further simplified by $\nabla l_\theta(X \cup \{x, y_i\}) \approx \nabla l_\theta(\{x, y_i\})$ since the loss function converged before the new query and the gradient of the norm of the loss function based on the last training set is almost zero. Empirical studies have shown the effectiveness of this approach, but it can be computationally demanding when dealing with large feature spaces and label sets.

Similar strategies, such as expected error reduction and variance reduction, are also applicable to quantum ML, but it is essential to conduct theoretical analyses on the loss function. In the case of expected error reduction, which is another decision theoretic approach, the goal is to estimate how much the generalization error of the model is likely to be reduced. This strategy calculates the expected future error of a model taught on the training set that includes each candidate individually and selects the sample with the least risk from the remaining unlabeled pool. It is important to note that directly minimizing the expectation of a quantum loss function is even more computationally expensive than in the classical case. Variance reduction, on the other hand, reduces generalization error by minimizing output variance, and in some cases, it can be solved analytically with a closed-form solution. Interestingly, the variance reduction method has already found applications in the field of material science, as we discussed earlier. These strategies offer valuable avenues for exploration and advancement in the field of quantum ML, but their implementation requires careful consideration of computational costs, theoretical analysis, and the specific requirements of the learning task at hand.

**Density-weighted methods**

The concepts of estimated error and variance reduction in AL offer a distinct approach by focusing on the entire input space rather than individual instances. This characteristic sets them apart from simpler query strategies like uncertainty sampling, query by committee, and expected gradient length, which are more susceptible to querying outliers. In certain cases, the least certain sample may lie on the classification boundary but may not be the most informative in terms of the overall data distribution. Query by committee and expected gradient length strategies may also exhibit similar behavior, spending time querying potential outliers due to their controversial nature or their expected impact on the model. Although estimated error and variance reduction strategies implicitly address these issues, they can be further improved by considering the density information during the query process. The key idea is that informative instances should not only be uncertain but also representative of the underlying distribution. By incorporating density information, a more comprehensive and effective AL strategy can be developed. By multiplying an additional term with density information to the basic query strategy $\phi(x)$, we write down the following criteria

$$x = \mathrm{argmax}_x \phi(x) \left[ \frac{1}{U} \sum sim(x, x^{(u)}) \right]^\beta,$$

which characterizes the average similarity of the sample to all other samples in the unlabeled pool, while its importance is controlled by $\beta$.

The concept of density-weighted sampling in AL provides inspiration for incorporating physical insights, particularly those related to quantum information. When employing a variational quantum circuit for post-processing the outcomes of quantum experiments, one can consider the distance between experimental parameters producing the samples in the parameter space to guide the density-weighted method. Similarly, distances between classical unlabeled samples can be calculated prior to encoding them in a quantum system for the same purpose. However, in the realm of quantum mechanics, the infidelity between quantum states can be leveraged as a measure of information density, leading to a variant of quantum AL. Specifically, our main idea revolves around the notion that unlabeled quantum samples with higher average infidelity to the training set are more likely to introduce additional information when measured and labeled. By incorporating this infidelity-based density measure, we can enhance the effectiveness of quantum AL strategies. Thus, we have

$$x_{IF} = \mathrm{argmax}_x \phi(x) \left[ \frac{1}{X} \sum_i (1 - |\langle x | x_i \in X \rangle|^2) \right]^\beta,$$

which is in the language of quantum information theory.

## Conclusion

In summary, this review has aimed to provide a comprehensive and accessible introduction to the concept of AL for physicists who may be interested in this field. We have highlighted the diverse range of applications of AL in various areas of physical science, spanning from quantum information to high-energy physics, nuclear physics, plasma physics, material science, and many-body physics. By showcasing these applications, we have demonstrated the wide-reaching potential of AL in advancing scientific research. Furthermore, we have presented our own perspective on the exciting intersection of AL and quantum ML based on our experience in both topics, reflected in two papers already published [Ding2020, Ding2022] and ongoing research. In this context, where the cost of labeling becomes increasingly expensive, we believe that AL offers a promising approach to mitigate this challenge. We have proposed alternative strategies that specifically leverage the advantages of AL in the training of quantum neural networks. By combining the principles of AL with the power of quantum ML, researchers can harness the benefits of both fields and unlock new possibilities for tackling complex scientific problems. We hope that this review serves as a valuable resource and inspiration for physicists interested in exploring the potential of AL and its fusion with quantum ML. As this field continues to evolve, we anticipate exciting advancements and fruitful collaborations between the realms of AL and quantum physics.

## Acknowledgments

This work was financially supported by EU FET Open Grant EPIQUS (899368); HORIZON-CL4-2022-QUANTUM-01-SGA project 101113946 OpenSuperQPlus100 of the EU Flagship on Quantum Technologies; the Basque Government through Grant No. IT1470-22; the project grant PID2021-126273NB-I00 funded by MCIN/AEI/10.13039/501100011033 and by "ERDFA way of making Europe" and "ERDF Invest in your Future"; Nanoscale NMR and complex systems (PID2021-126694NB-C21) and QUANTEK project (Grant No. KK-2021/00070); the Valencian Government Grant with Reference Number CIAICO/2021/184; the Spanish Ministry of Economic Affairs and Digital Transformation through the QUANTUM ENIA project call—Quantum Spain project, and the European Union through the Recovery, Transformation and Resilience Plan—NextGenerationEU within the framework of the Digital Spain 2025 Agenda; NSFC (12075145); STCSM (Grant No. 2019SHZDZX01-ZX04). X. C. acknowledges 'Ayudas para contratos Ramón y Cajal'—2015–2020 (RYC-2017-22482). Y. D. acknowledeges the sponsorship from Chinese Physical Society and MindSpore Quantum for quantum machine learning.

# References

[Biamonte2017] J. Biamonte, P. Wittek, N. Pancotti, P. Rebentrost, N. Wiebe, S. Lloyd, Quantum machine learning, Nature 549, 195–202 (2017).

[Caron2019] S. Caron, T. Heskes, S. Otten, and B. Stienen, Constaining the parameters of high-dimensional models with active learning, Eur. Phys. J. C 79, 944 (2019).

[Casares2020] P. A. M. Casares and M. A. Martin-Delgado, A quantum active learning algorithm for sampling against adversarial attacks, New J. Phys. 22, 073026 (2020).

[Diaw2020] A. Diaw, K. Barros, J. Haack, C. Junghans, B. Keenan, Y. W. Li, D. Livescu, N. Lubbers, M. McKerns, R. S. Pavel, D. Rosenberger, I. Sagert, and T. C. Germann, Multiscale simulation of plasma flows using active learning, Phys. Rev. E 102, 023310 (2020).

[Ding2020] Y. Ding, J. D. Martín-Guerrero, M. Sanz, R. Magdalena-Benedicto, X. Chen and E. Solano, Retrieving Quantum Information with Active Learning, Phys. Rev. Lett. 124, 140504 (2020).

[Ding2022] Y. Ding, J. D. Martín-Guerrero, Y. Song, R. Magdalena-Benedicto, and X. Chen, Active learning for the optimal design of multinomial classification in physics, Phys. Rev. Research 4, 013213 (2022).

[Ding2023] C. Ding, X.-Y. Xu, Y.-F. Niu, S. Zhang, W.-S. Bao, and H.-L. Huang, Active Learning on a Programmable Photonic Quantum Processor, Quantum Sci. Technol. 8, 035030 (2023)

[Krenn2016] M. Krenn, M. Malik, R. Fickler, R. Lapkiewicz, and A. Zeilinger, Automated Search for new Quantum Experiments, Phys. Rev. Lett. 116, 090405 (2016).

[Lange2023] H. Lange, M. Kebrič, M. Buser, U. Schollwöck, F. Grusdt, and A. Bohrdt, Adaptive Quantum State Tomography with Active Learning, arXiv: 2203.15719

[Lysogorskiy2023] Y. Lysogorskiy, A. Bochkarev, M. Mrovec, and R. Drautz, Phys. Rev. Materials 7, 043801 (2023).

[Melnikov2018] A. A. Melnikov, H. P. Nautrup, M. Krenn, V. Dunjko, M. Tiersch, A. Zeilinger, and H. J. Briegel, Active learning machine learns to create new quantum experiments, Proc. Natl. Acad. Sci. U.S.A. 115, 1221 (2018).

[Mroczek2023] D. Mroczek, M. Hjorth-Jensen, J. Noronha-Hostler, P. Parotto, C. Ratti, and R. Vilata, Mapping out the thermodynamic stability of a QCD equation of state with a critical point using active learning, Phys. Rev. C 107, 054911 (2023).

[Paparo2014] G. D. Paparo, V. Dunjko, A. Makmal, M. A. Martin-Delgado, and H. J. Briegel, Quantum Speedup for Active Learning Agents, Phys. Rev. X 4, 031002 (2014).

[Rocamonde2022] J. Rocamonde, L. Corpe, G. Zilgalvis, M. Avramidou, and J. Butterworth, Picking the low-hanging fruit: testing new physics at scale with active learning, SciPost Phys. 13, 002 (2022).

[Sahinovic2021] A. Sahinovic and B. Geisler, Active learning and element-embedding approach in neural networks for infinite-layer versus perovskite oxides, Phys. Rev. Materials 3, L042002 (2021).

[Settles2009] B. Settles, Active learning literature survey (2009).

[Seung1992] H. S. Seung, M. Opper, H. Sompolinsky. Query by committee, Proc. of the ACM Workshop on Computational Learning Theory, Pittsburgh (PA, USA) (1992).

[Sverchkov2017] Y. Sverchkov, M. Craven, A review of active learning approahes to experimental design for uncovering biological networks, PLoS Comput. Biol. 13, e1005466 (2017).

[Tian2021] Y. Tian, D. Xue, R. Yuan, Y. Zhou, X. Ding, J. Sun, and T. Lookman, Efficient estimation of material property curves and surfaces via active learning, Phys. Rev. Materials 5, 013802 (2021).

[Tuia2011] D. Tuia, M. Volpi, L. Copa, M. Kanevski, and J. Muñoz-Marí, A survey of active learning algorithms for supervised remote sensing image classification, IEEE J. Sel. Top. Signal Process. 5, 606 (2011).

[Wu2020] Y. Wu, Z. Meng, K. Wen, C. Mi, J. Zhang, and H. Zhai, Active Learning Approach to Optimization of Experimental Control, Chinese Phys. Lett. 37 103201 (2020).

[Yao2020] J. Yao, Y. Wu, J. Koo, B. Yan, and H. Zhai, Active learning algorithm for computational physics, Phys. Rev. Research 2, 013287 (2020).

[Zhang2019] L. Zhang, D.-Y. Lin, H. Wang, R. Car, and W. E, Active learning of uniformly accurate interatomic potentials for materials simulation, Phys. Rev. Materials 3, 023804 (2019).

[Zhu2010] J. Zhu, H. Wang, B. K. Tsou, M. Ma, Active Learning With Sampling by Uncertainty and Density for Data Annotations, IEEE Trans. Audio. Speech. Lang. Process., 18(6), 1323-1331 (2010).

[Zhu2022] R. Zhu, C. Pike-Burke, and F. Mintert, Active Learning for Quantum Mechanical Measurements, arXiv: 2212.07513.